Zinc-blende and wurtzite GaAs quantum dots in nanowires studied using hydrostatic pressure

Shuang Yang, Kun Ding, Xiuming Dou, Xuefei Wu, Ying Yu, Haiqiao Ni, Zhichuan Niu, Desheng Jiang, Shu-Shen Li, Jun-Wei Luo[*] & Baoquan Sun[*]

State Key Laboratory for Superlattices and Microstructures, Institute of Semiconductors, Chinese Academy of Sciences, Beijing 100083, China

We report both zinc-blende (ZB) and wurtzite (WZ) crystal phase self-assembled GaAs quantum dots (QDs) embedding in a single GaAs/AlGaAs core-shell nanowires (NWs). Optical transitions and single-photon characteristics of both kinds of QDs have been investigated by measuring photoluminescence (PL) and time-resolved PL spectra upon application of hydrostatic pressure. We find that the ZB QDs are of direct band gap transition with short recombination lifetime (~1 ns) and higher pressure coefficient (75-100 meV/GPa). On the contrary, the WZ QDs undergo a direct-to-pseudodirect bandgap transition as a result of quantum confinement effect, with remarkably longer exciton lifetime (4.5-74.5 ns) and smaller pressure coefficient (28-53 meV/GPa). These fundamentally physical properties are further examined by performing state-of-the-art atomistic pseudopotential calculations.

*e-mail: bqsun@semi.ac.cn; jwluo@semi.ac.cn

Quantum dots (QDs) with a direct bandgap, which are embedded in photonic nanowires (NWs), are recognized as being an excellent candidate for bright single-photon source, such as zinc-blende (ZB) GaAs and InAs QDs.[1,2] However, for single-photon quantum storage and detection,[3-5] semiconductor materials with an indirect bandgap are usually chosen due to their long exciton lifetime. It is known that GaAs crystal is typically of cubic ZB phase. The hexagonal wurtzite (WZ) crystal phase is not available in GaAs bulk and thin film under common synthetic conditions, but was recently observed in GaAs NWs.[6-9] In WZ GaAs, the $\overline{\Gamma}_{7c}$ ($\overline{\Gamma}_{8c}$) band is a bright (dark) conduction band, corresponding to the $\Gamma_{6c}$ ($L_{6c}$) band in the ZB GaAs phase.[10] The hexagonal crystal field significantly reduces the energy difference between $\overline{\Gamma}_{7c}$ and $\overline{\Gamma}_{8c}$ conduction bands in bulk WZ phase,[11,12] leading to a debate regarding whether the bright or dark band is the lowest conduction band. For instance, the bright conduction band $\overline{\Gamma}_{7c}$ was claimed to be the lowest conduction band by some theoretical calculations and experimental measurements,[10-12] giving rise to a direct bandgap transition. However, other researchers argued for a pseudodirect bandgap, where the dark conduction band $\overline{\Gamma}_{8c}$ is the lowest conduction band.[13,14] In comparison with their NW counterpart, WZ single QDs have not been found in III-V non-nitride semiconductors. WZ GaAs QDs may have an indirect bandgap transition due to the spatial quantum confinement effects, which will further shift $\overline{\Gamma}_{7c}$ up relative to $\overline{\Gamma}_{8c}$. Therefore, it is of great interest to investigate WZ GaAs QDs by performing the experimental and theoretical investigations of the bandgap transition.

In this report, the optical transitions and single-photon characteristics of both WZ and ZB GaAs QDs embedded in NWs were investigated by photoluminescence (PL) and time-resolved PL spectra upon application of hydrostatic pressure. These two types of QDs show different emission energy, exciton lifetime, and pressure behaviors, thus suggesting ZB QDs as an excellent bright single-photon source and a direct-to-pseudodirect bandgap transition in WZ QDs. The fundamental physical properties are further examined by performing state-of-the-art atomistic pseudopotential calculations.

The investigated GaAs QDs in NWs with a core-shell GaAs/Al$_{0.7}$Ga$_{0.3}$As structure were grown by Molecular Beam Epitaxy (MBE) on a GaAs (001) substrate in a self-catalyzed vapor-liquid-solid mechanism. Specifically, in sequence, the NW consists of a ~200 nm GaAs core layer, a layer of ~50 nm Al$_{0.7}$Ga$_{0.3}$As shell, a GaAs QDs layer, a ~50 nm Al$_{0.7}$Ga$_{0.3}$As shell layer, and finally, a ~5 nm outer GaAs layer. A confocal microscopic spectral system was used for measuring the PL spectra, time-resolved PL spectra, and the second-order correlation function $g^{(2)}(\tau)$. The sample was mounted inside a helium-free cryostat at a temperature of 6 K. Excitation was performed with a continuous wave (CW) 454 nm semiconductor laser or a CW/pulse adjustable 640 nm semiconductor laser. The PL of the samples was collected through a 50 × microscope objective (NA: 0.5), dispersed with a 0.5 m monochromator and detected with a silicon charge-coupled detector (Si-CCD).

Silicon single-photon counting modules with a time resolution of 40 ps were used for time-resolved PL and the second-order correlation function $g^{(2)}(\tau)$ measurements.

To study the pressure behavior of the QD emission lines at low temperature, the NWs were transferred to a GaAs substrate and covered by a ~500 nm $SiO_2$ layer in case the NWs fall off during mechanical thinning. After being mechanically thinned to ~20 μm, the sample (approximately 100 × 100 μm$^2$) was placed into the diamond anvil cell (DAC) chamber. Figure 1 shows the schematics of the improved diamond anvil cell (DAC) device. The DAC pressure device was driven by a piezoelectric actuator through an external voltage source. With this device, we can study the behavior of single GaAs QDs in GaAs/AlGaAs NWs with continuously changing pressure at low temperature to avoid missing or confusing particular single QD PL signals. The condensed argon was used as the pressure-transmitting medium, and the ruby $R_1$ fluorescence line was used to calibrate the pressure. The calibrated temperature of the sample chamber is 18–20 K.[15]

We use the empirical pseudopotential method to solve the problem of multi-million atom QDs with discrete atoms that are located at specific positions, each carrying its own (screened) pseudopotential. This description forces upon us the correct atomically-resolved symmetry, thereby including automatic effects of shape, strain, alloy fluctuations, and wave function mixing such as inter-band mixing, intra-band mixing, and inter-valley mixing as well as the direct-to-indirect bandgap transition for GaAs QDs.[16]

Figure 2(a) shows a typical PL spectrum of a single NW with many emission lines distributed within the energy window from 1.72 to 1.96 eV, indicating that there is more than one QDs embedded in the NW. Time-resolved PL is then applied to characterize the dynamic excitonic behavior of individual QDs, and it is surprising to note that the excitonic dynamic behavior is rather different among various emission lines of QDs, even embedded in the same single NW. Figure 2(b) shows the exciton decay of two QDs with the lifetime values of 16.63 and 0.27 ns for the emission lines at 1.74 and 1.73 eV (the PL emission lines are not shown here), respectively. The exciton lifetimes of all measured 39 QDs spread over a wide range from 0.19 to 74.46 ns and are given in Fig. 2(c) as a function of exciton energy. The exciton energies of QDs with long exciton lifetimes fall in a narrow range approximately 1.7 eV; however, the energies of QDs with short lifetimes disperse in a large range from 1.6 to 1.9 eV. This result suggests that there are two different optical transition mechanisms among the various QDs embedded in the NWs. One type holds long radiative lifetime and exciton energies that are always approximately 1.7 eV, and the other type has a short lifetime but its exciton energies are spread over a large range. The short radiative lifetime is typically within the range of direct bandgap emissions of self-assembled QDs.[17] We could assign this emission to direct bandgap transition, whereas the longer lifetime indicates that the corresponding QDs possess a fairly different exciton recombination mechanism, which will be discussed in detail later.

To unravel the origin of the observed diverse exciton lifetimes of the GaAs QDs, we investigate the QD PL upon application of hydrostatic pressure. Figure 3(a) shows the pressure response of 13 emission lines of GaAs QDs, which as expected are all linearly blue-shifted with increasing hydrostatic pressure. The obtained pressure coefficients (PCs) of individual emission lines together with their corresponding radiative lifetimes at ambient pressure are summarized in Fig. 3(b), where all measured QDs are clearly separated into two groups: one group has large PCs approximately 75–100 meV/GPa associating with short exciton lifetimes (~1 ns), and the other group has small PCs approximately 28–53 meV/GPa accompanying long exciton lifetimes (4.5-74.46 ns). This result demonstrates that those QDs embedded in a single NW have two different optical transition mechanisms. To assign the two transition mechanisms, we first briefly review the bulk ZB GaAs optical transitions, where the PCs are ~105 meV/GPa, ~55 meV/GPa and ~-15 meV/GPa for the direct bandgap $E_g^\Gamma$ occurs at the Γ-point, the Γ-L indirect bandgap $E_g^L$ and the Γ-X indirect bandgap $E_g^X$, respectively.[18] It should be noted that the bandgap PC of self-assembled QDs is usually a little smaller than the corresponding parent bulk semiconductors[19] due to the influences of built-in strain, energy barrier height, effective mass and QD size.[20,21] There is no doubt that the observed QD emissions with short radiative lifetime and large PC are direct bandgap transitions. The QD emissions with long lifetime and small positive PCs, however, should most likely come from the QD indirect bandgap transition, which originally corresponds to the indirect bandgap Γ-L transition in bulk GaAs.

In ZB GaAs bulk, the direct bandgap $E_g^\Gamma$ is 1.519 eV and the Γ-L indirect gap $E_g^L$ is 1.98 eV.[22] The large energy separation (460 meV) between Γ-valley and L-valley presents difficulties in engineering ZB GaAs to be indirect upon application of a small perturbation. However, in bulk WZ GaAs, the Γ-L separation is significantly reduced due to the strong coupling between L-derived states induced by a hexagonal crystal field.[11,12] One of four ZB L-valley folds to the WZ $\overline{\Gamma}$-point and forms a $\overline{\Gamma}_{8c}(L_{6c})$ state, a dark state transition to the valence band in terms of the selection rules, which is above $\overline{\Gamma}_{7c}(\Gamma_{6c})$ state in energy. The $\overline{\Gamma}_{8c}(L_{6c})$ state is pushed down and is closer to $\overline{\Gamma}_{7c}(\Gamma_{6c})$ by an interband coupling. Consequently, in WZ GaAs bulk, even a small perturbation (*e.g.*, a compressive uniaxial stress) could lead to a direct-to-pseudodirect bandgap transition by reversing the band order of $\overline{\Gamma}_{7c}(\Gamma_{6c})$ and $\overline{\Gamma}_{8c}(L_{6c})$ as reported in WZ GaAs NWs.[11] Such direct-to-pseudodirect bandgap transition can also be driven by spatial confinement such as in QDs due to a much heavier effective mass of $\overline{\Gamma}_{8c}(L_{6c})$ than that of $\overline{\Gamma}_{7c}(\Gamma_{6c})$. In previous work, we already demonstrated such a direct-to-pseudodirect bandgap transition occurring in ZB GaAs QDs, which has an extremely small dot size of approximately 3.3 nm and a large bandgap of approximately 2.3 eV.[16] Such a transition is expected to occur at a larger dot size and smaller bandgap in the WZ phase because of a much smaller Γ-L energy separation. We therefore assign the group of QDs with features of smaller bandgap PC, longer radiative lifetime and bandgap approximately 1.7 eV to WZ phase GaAs QDs. In fact, the formation of both ZB and WZ QDs within a single NW may be because the NW has ZB and WZ segments stacking alternately along the NW axis, which is revealed by the high-resolution transmission electron microscopy (HRTEM). Figure 3(c) shows that both ZB and WZ crystal phases of the NW are found along the NW axis as marked by red circles, respectively. The QD layer is a sheath of the NW in terms of our growth recipe, and it is natural to expect that the QDs located within the ZB NW segments must also own a ZB phase and QDs within the WZ NW segments own a WZ phase. This result strongly supports our previous assignments. To the best of our knowledge, this is the first report on the synthetic and observation of WZ QDs made of non-nitride III-V semiconductors.

To understand the quantum confinement effect on the direct-to-pseudodirect transition in WZ GaAs QDs, we have performed state-of-the-art atomistic calculations for GaAs QDs embedded in an $Al_{0.6}Ga_{0.4}As$ matrix with both ZB and WZ crystal phases. The calculated WZ bandgap transition has an optical transition matrix element (in atomic units) of $\mu = |\langle \Gamma_{9v} | p | \Gamma_{7c} \rangle|^2 = 0.304$ close to the bandgap transition μ=0.354 in the ZB phase, indicating that the WZ GaAs is direct bandgap and is in agreement with the recent experimental measurement.[11] The slightly high lying $\overline{\Gamma}_{8c}(L_{6c})$ state is a dark transition to valence $\overline{\Gamma}_{9v}(\Gamma_{8v})$ state with predicted $\mu = |\langle \Gamma_{9v} | p | \Gamma_{8c} \rangle|^2 = 1.97 \times 10^{-5}$. Figure 4 shows the calculated absorption spectra of both ZB and WZ GaAs QDs embedded in an $Al_{0.6}Ga_{0.4}As$ matrix. Both QDs have the same lens shape and size, 12 nm base diameter and 5.5 nm dot height, but different ZB and WZ crystal phases to model the QDs in NWs. The bandgap transition of the ZB GaAs QD is predicted to be bright with an optical transition matrix $\mu = 0.253$, as shown in Fig. 4(a), whereas the predicted bandgap transition of the WZ QD is four orders of magnitude weaker with $\mu = 4.05 \times 10^{-5}$ and the lowest bright transition with $\mu = 0.209$ is approximately 0.1 eV higher than the bandgap, as shown in Fig.4(b). The dark bandgap transition of WZ QD is responsible for bulk $\overline{\Gamma}_{8c}(L_{6c}) - \overline{\Gamma}_{9v}(\Gamma_{8v})$ transition and the highly lying bright transition for bandgap transition in bulk WZ GaAs. Hence, we predict a direct-to-pseudodirect bandgap transition in WZ GaAs QD as a result of quantum confinement. These theoretical results strongly support our assignment of the experimentally observed weak emission with a small bandgap PC and long exciton lifetime to be the pseudodirect bandgap transitions of WZ GaAs QDs, and the strong emission with a large bandgap PC and short exciton lifetime to be the direct bandgap transitions of ZB GaAs QDs.

Figures 5(a)–(c) and 5(d)–(f) show the PL spectra, PL integrated counts per second as a function of the excitation power, and second-order correlation function $g^{(2)}(\tau)$ for typical short and long lifetime QDs, respectively, under an excitation of the CW 640 nm semiconductor laser at 6 K. With increasing excitation power, the energy of emission lines are not shifted, which rules out the possibility of type-Ⅱ transition between ZB and WZ GaAs.[23,24] It is found that both QDs show high single-photon purity with $g^{(2)}(\tau=0) \sim 0.1$ near the saturation point of the excitation power, as seen in Fig. 5(c) and 5(f). The photon counting rates at the saturation point of the excitation power for typical short and long lifetime QDs are approximately 1 MHz and 0.03 MHz, respectively. The former QDs responsible for ZB QDs can be applied to ideal single-photon sources,[25] whereas the latter one responsible for WZ QDs is more suitable for use as a single-photon quantum storage device or a photodetector.

To summarize, metastable WZ crystal phase, enriching the properties to non-nitride III-V bulk semiconductors, has been reported to be stabilized in NWs but not in QDs at ambient conditions. Here, we report for the first time that GaAs WZ phase in self-assembled QDs embedded within GaAs/AlGaAs core-shell NWs. We experimentally and theoretically investigate the optical transitions and single-photon characteristics of both ZB and WZ GaAs QDs in NWs. We demonstrate that the self-assembled WZ GaAs QDs in the sheath of the WZ NW segments undergo a direct-to-pseudodirect bandgap transition and are suitable for applications as photodetectors, whereas ZB GaAs QDs in the ZB NW segments exhibit excellent single-photon emission features. Given that ZB/WZ segments can be accurately engineered,[26,27] it is possible to precisely position the direct bandgap GaAs ZB QDs and pseudodirect bandgap GaAs WZ QDs within a single NW. These results may provide a way to integrated single-photon emitters and detectors in a single nanowire for advanced photonics and sensing applications.

The authors acknowledge support from the National Key Basic Research Program of China (Grant Nos. 2013CB922304, 2013CB933304), the National Natural Science Foundation of China (Grant Nos. 11474275, 61474116), the Strategic Priority Research Program (B) of the Chinese Academy of Sciences (Grant No. XDB01010200), and the National Young 1000 Talents Plan.

Figures

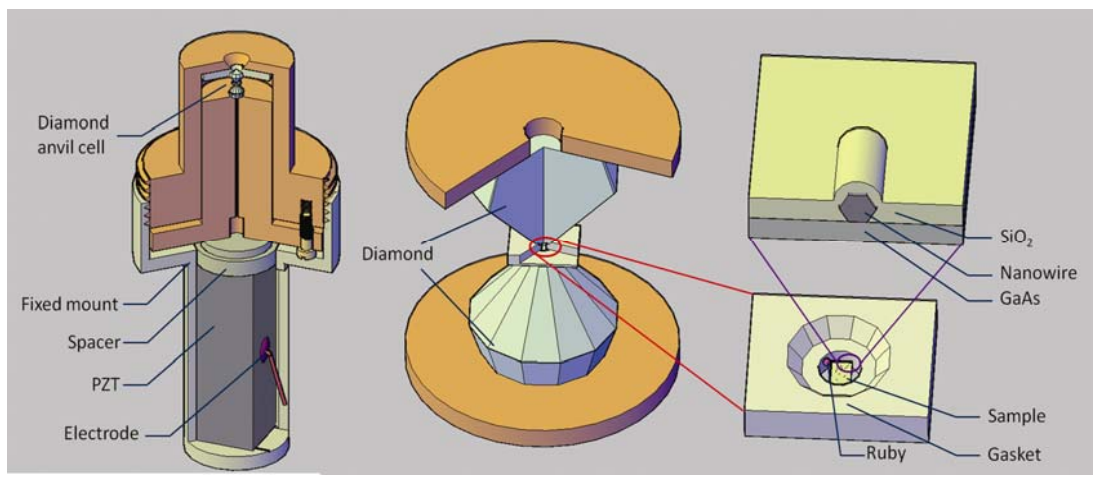

FIG. 1. (color online) Schematics of the applied pressure device (left), diamond anvil cell (middle), and the amplified diagram of the sample chamber with QDs in NW sample and the ruby (right).

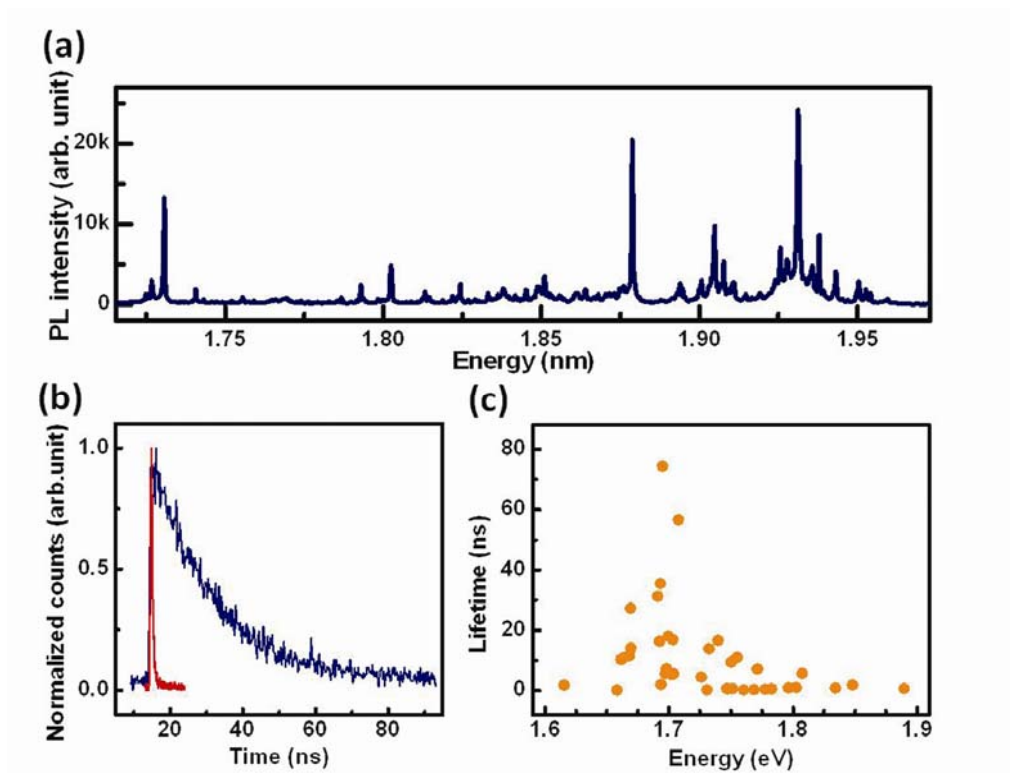

FIG. 2. (color online) (a) The PL of a single nanowire at 6 K. (b) The time-resolved spectra of two typical QD emission lines. The obtained lifetimes are 16.63 and 0.27 ns for blue and red lines, respectively. (c) Lifetime versus photon energy of 39 QD emission lines.

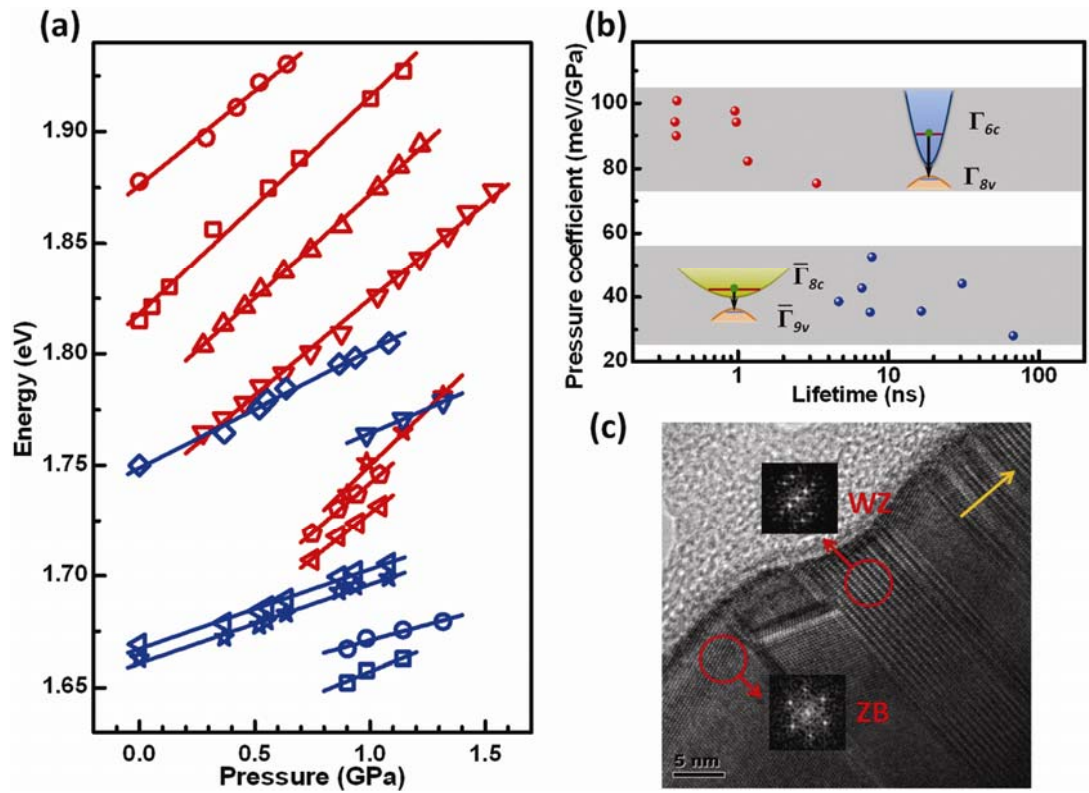

FIG. 3. (color online) (a) PL peak energy of GaAs QDs as a function of pressure. There are two types of pressure responses of QDs emission. (b) Relation between the PL emission lifetime and pressure coefficient. The red (blue) points correspond to the short (long) lifetime QDs. Inset: the schematic diagrams show two distinct bandgap transitions. (c) A HRTEM image and diffraction patterns of the NW. The yellow arrow indicates the NW growth direction. The two different regions show ZB and WZ crystal structures, respectively.

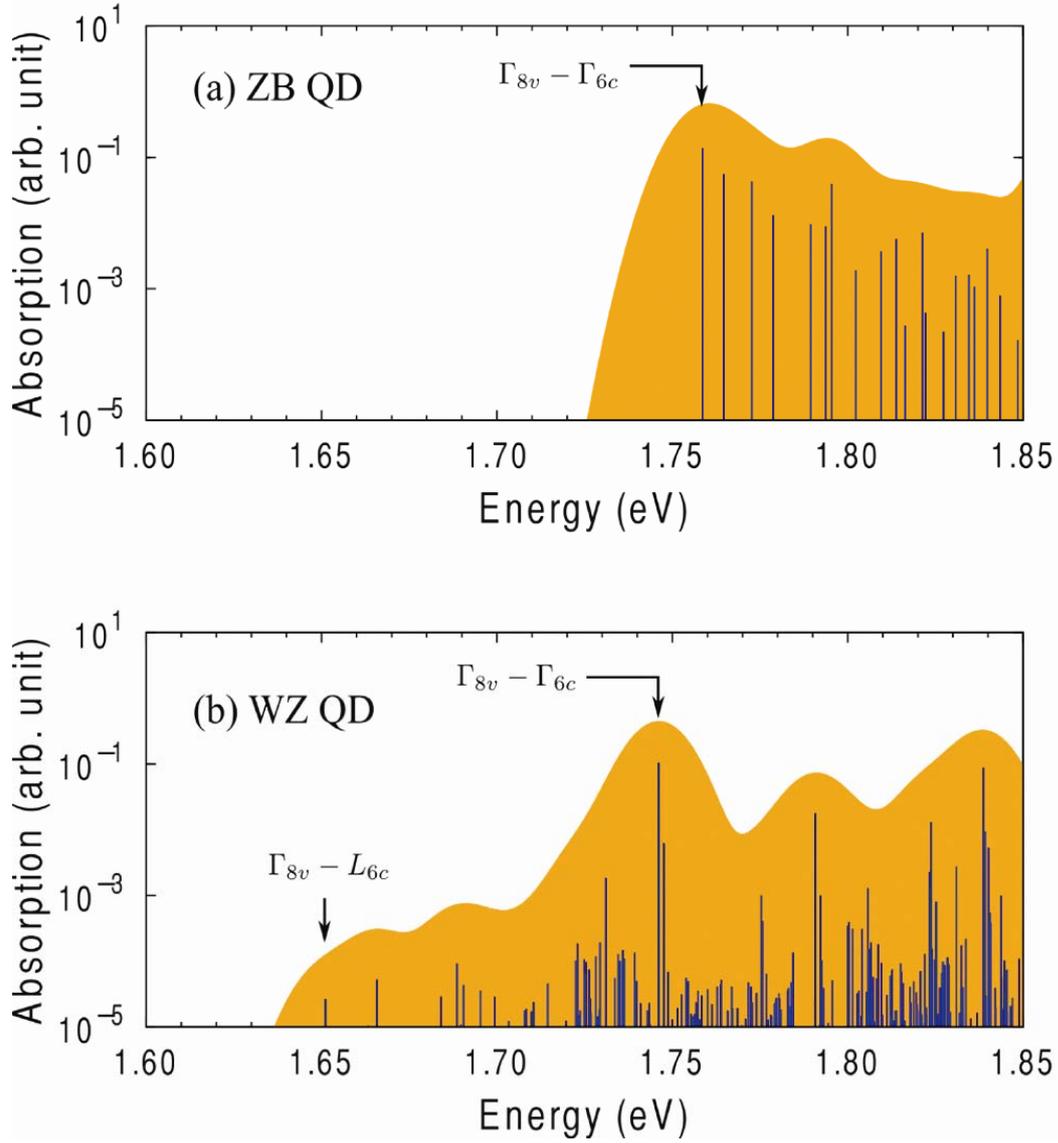

FIG. 4. (color online) The absorption spectra of ZB (a) and WZ (b) GaAs QDs, respectively. Both QDs have the same lens shape with a 12 nm base diameter and 5.5 nm dot height and were embedded in an $Al_{0.6}Ga_{0.4}As$ matrix to model the QDs in NWs. The vertical lines represent the magnitude of the optical transition matrix elements. The arrows indicate the optical transitions of the QDs corresponding to the bulk ZB or WZ GaAs.

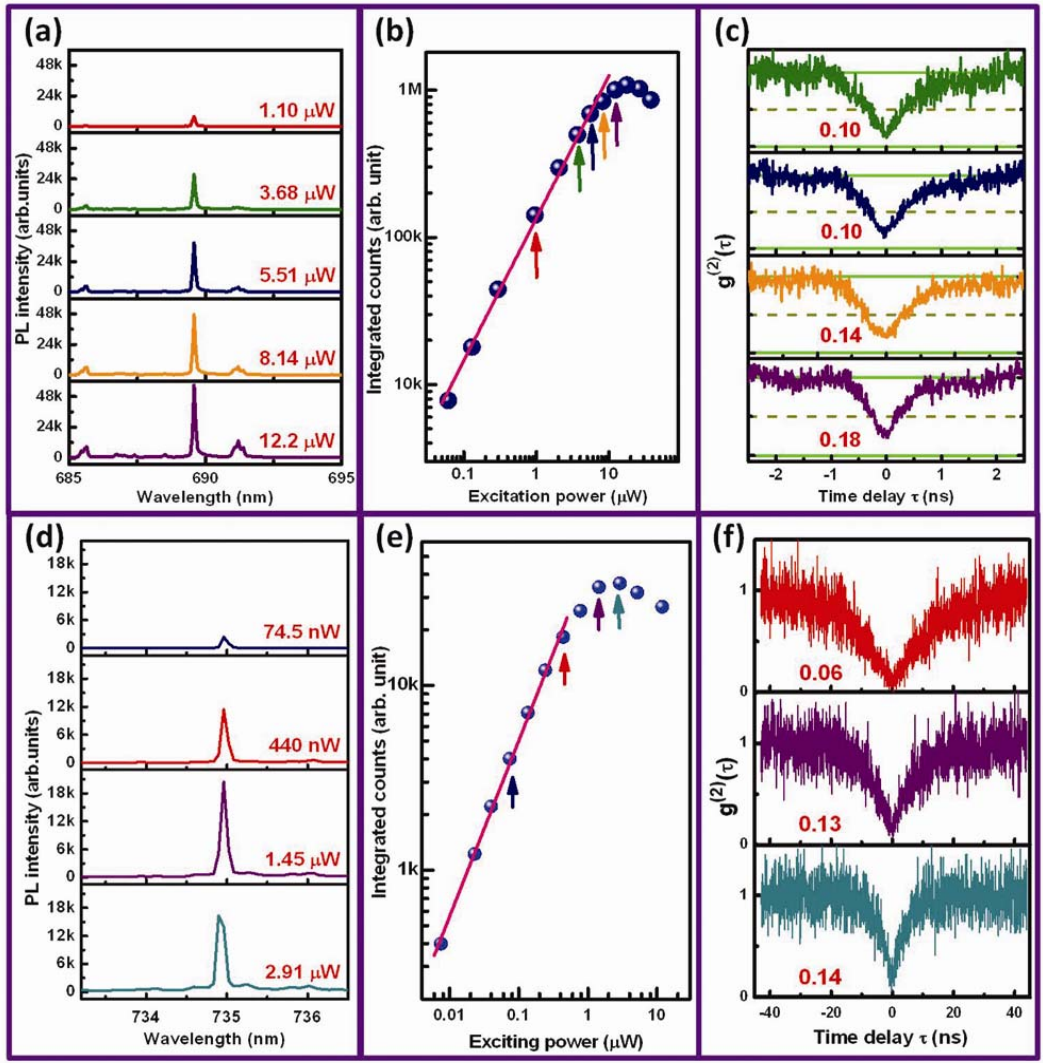

FIG. 5.(color online) PL spectra observed under different excitation powers, integrated counts per second as a function of excitation power and second-order correlation function curves for a typical short (a)–(c) and long (d)–(f) lifetime QD single-photon emission lines, respectively, where the same color curves correspond to the same colored PL emission lines.